\documentclass[aps,pre,reprint]{revtex4-1}
\usepackage{graphicx}
\usepackage[utf8]{inputenc}
\usepackage{amsmath}
\usepackage{amsfonts}
\usepackage{amssymb}
\usepackage{upgreek}

\begin{document}

\title{On the beneficial role of noise in resistive switching}

\author{G.~A.~Patterson}
\email[]{german@df.uba.ar}
\affiliation{Instituto Tecnol\'ogico de Buenos Aires, Av. E. Madero 399 (C1106ACD), C.A.B.A., Argentina}
\author{P.~I.~Fierens}
\affiliation{Instituto Tecnol\'ogico de Buenos Aires}
\affiliation{Consejo Nacional de Investigaciones Cient\'ificas y T\'ecnicas, Argentina}
\author{D.~F.~Grosz}
\affiliation{Instituto Tecnol\'ogico de Buenos Aires}
\affiliation{Consejo Nacional de Investigaciones Cient\'ificas y T\'ecnicas, Argentina}

\date{\today}

\begin{abstract}
We study the effect of external noise on resistive switching. Experimental results on a manganite sample are presented showing that there is an optimal noise amplitude that maximizes the contrast between high and low resistive states. By means of numerical simulations, we study the causes underlying the observed behavior. We find that experimental results can be related to general characteristics of the equations governing the system dynamics.
\end{abstract}


\maketitle

\section{Introduction}

The phenomenon of resistive switching (RS) is usually associated with a passive device called \textit{memristor}, first proposed in 1971 by L. Chua \cite{Chua.IEEETransCircuitTheory.1971}. Since then many mechanisms were proposed to explain the behavior of resistive switching materials (see, e.g., Refs. \cite{Sawa.MaterialsToday.2008, Strukov.Nature.2008, Rozenberg.PhysRevB.2010} and references therein). One of the applications of RS devices is as part of resistive random access memories (ReRAMs), promising candidates to succeed current storage technologies \cite{Sawa.MaterialsToday.2008}. The high level of integration required in these and other applications leads naturally to the study of the behavior of resistive switching under noisy conditions. 

In this direction, Stotland and Di Ventra \cite{Stotland.PRE.2012} showed that internal noise helps to increase the contrast ratio between low and high resistive states. The results in Ref.~\cite{Stotland.PRE.2012} were based on simulations by means of a model of RS put forth by Strukov \textit{et al}. \cite{Strukov.Nature.2008}. In Ref. \cite{Patterson.ICAND.2012}, however, it was shown that the presence of external noise (i.e., noise added to the input signal) is only detrimental under Strukov's model. Nevertheless, Patterson \textit{et al}. \cite{Patterson.PhysRevE.87} showed an experiment in which the addition of external noise had a positive effect in a manganite sample (see, e.g., \cite{Ghenzi.JournalAppliedPhysics.2010,Ghenzi.JournalAppliedPhysics.2012,Gomez-Marlasca.APL.2011} and references therein).

In this work we extend the results of Ref.~\cite{Patterson.PhysRevE.87} in two directions. On the one hand, in Section \ref{sec:exp_res} we present the results of a large number of experiments on the same type of manganite in order to study its behavior under varying noise conditions. In particular, we analyze a quality factor $Q$ which not only takes into account the contrast between high and low resistive states, but also their variance. On the other hand, in Section \ref{sec:sim_res} we study the influence of noise by means of simulations. 

In 1976, Chua and Kang \cite{Chua.IEEEProc.1976} defined a \textit{memristive system} as a nonlinear dynamical system described by 
\begin{eqnarray}
v(t) = R(x,i)i(t)\ ,\label{eq:ohm} \\
\frac{dx}{dt} = f(x,i)\ ,\label{eq:memristor_state}
\end{eqnarray}
where $v(t)$ is the applied voltage, $i(t)$ is the current, $R$ the resistance, $x$ a set of state variables, and $f(x,i)$ a nonlinear function that governs the evolution of $x$. In Ref.~\cite{Patterson.PhysRevE.87} a good agreement is observed between experiment and simulations based on the function $f(x,i)$ introduced in Refs.~\cite{Pickett.JournalofAppliedPhysics.2009,Kvatinsky.IEEETransactionsonCircuitsandSystemsI.2013}. Therefore, our simulations are based on this particular model. We also analyze the general characteristics of a function $f(x,i)$ which lead to results compatible with experiments.

Finally, in Section \ref{sec:concl} we present a summary of the main results and conclusions.

\section{\label{sec:exp_res}Experimental results}

\begin{figure}
\begin{center}
\includegraphics[width=1\columnwidth]{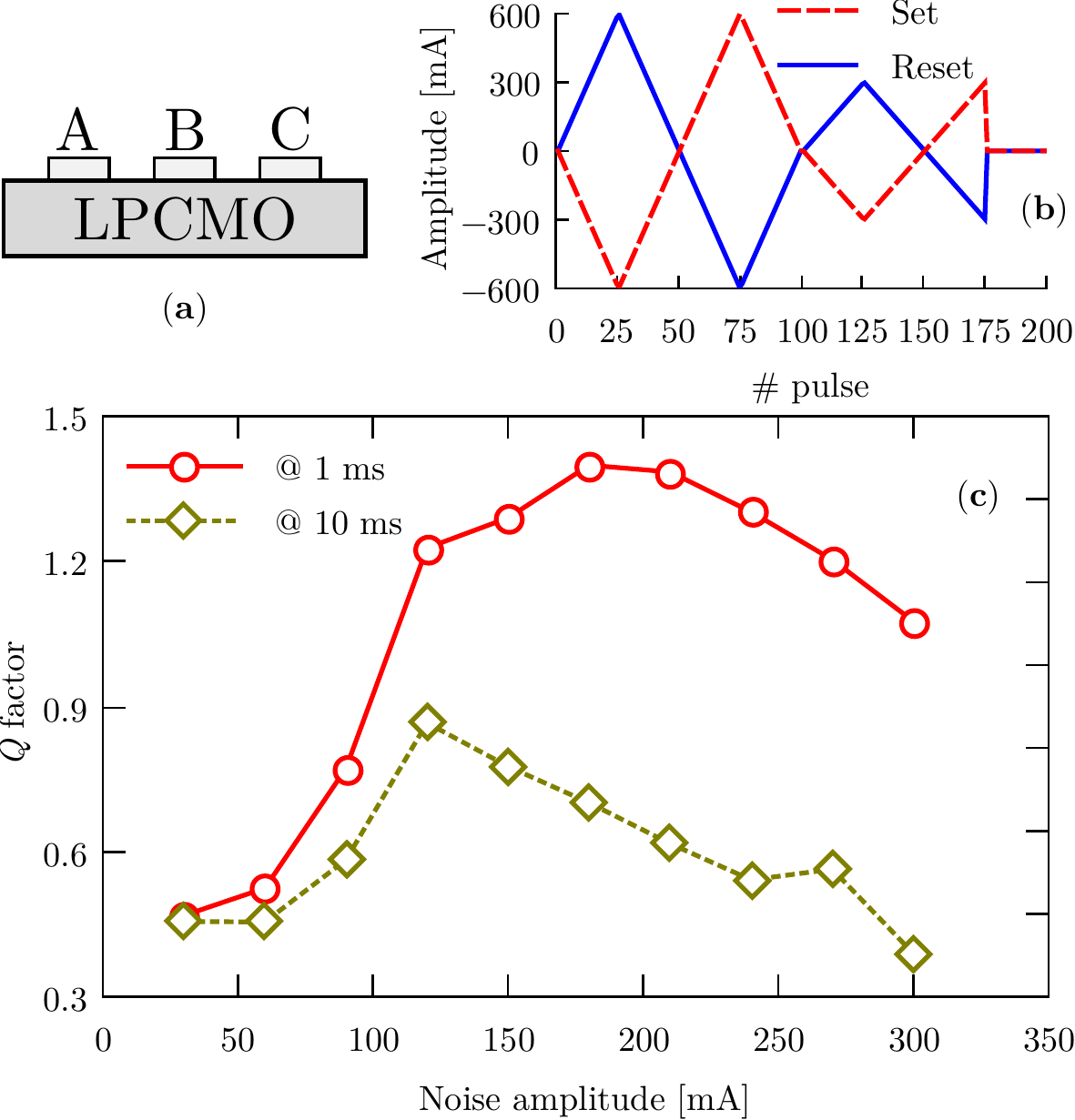}
\caption{(Color online). (a) sample connection scheme. (b) amplitude of current pulses. Noise is turned on after pulse \# 100. (c) quality factor as a function of noise amplitude. Results are shown for 1 and 10 ms after the input signal is turned off.}
\label{fig:Q_exp}
\end{center}
\end{figure}

Experiments were conducted on a polycrystalline sample of La$_{0.325}$Pr$_{0.300}$Ca$_{0.375}$MnO$_{3}$ (LPCMO). Current pulses were applied through two hand-painted silver electrodes. The connection scheme is shown in Fig.~\ref{fig:Q_exp}a. A third electrode was used to perform three-terminal measurements. Driving pulses were applied through contacts A and C, and the nonvolatile resistance was measured between contacts A-B and B-C due to the resistive switching phenomenon in this kind of compounds taking place near the contact surfaces \cite{Acha.PhysicaB.2009, Rozenberg.PhysRevB.2010}.

Experiments were performed as follows. Current pulses with a duration of 1 ms were applied in $\sim 1.5 \text{ s}$ intervals. Pulse amplitude was  varied as shown in Fig.~\ref{fig:Q_exp}b. An initial deterministic hysteresis switching loop was adopted to assure the best reproducible characteristics, erasing the pulsing history \cite{Gomez-marlasca.AppliedPhysicsLetters.2011}. Resistance measurements were performed by applying a small bias current (1 mA) during 1 ms immediately before the driving pulses.

Noise samples were generated at sample rate of 100 k Samples/s with a measured effective bandwidth of $\sim 75$ kHz. Noise amplitude (standard deviation) ranged from 30 to 300 mA. We measured the nonvolatile resistance after applying the set/reset protocol, and computed a quality factor $Q$ for each noise amplitude at time $t$ defined as
\begin{equation}
Q(t) = \frac{\left\langle R_{\text{off}}(t) \right\rangle -\left\langle R_{\text{on}}(t) \right\rangle }{\sigma_{\text{off}}(t) + \sigma_{\text{on}}(t)} \ ,
\label{eq:Q}
\end{equation}
where $\left\langle \cdot \right\rangle$ stands for the average over 100 noise realizations for the set ($R_{\text{on}}$) and reset ($R_{\text{off}}$) protocol, and $\sigma$ is the ensemble standard deviation. The $Q$ factor measures the contrast between high and low states and is commonly used in communication theory (see, e.g., \cite{Agrawal}). A higher $Q$ factor is thus associated with a smaller probability of error of discernment between the two states. 

Results are presented in Fig.~\ref{fig:Q_exp}c, showing a non-monotonic behavior of the $Q$ factor as a function of noise amplitude for two different elapsed times. There is a certain amount of noise that maximizes the quality factor in all considered cases despite the poor signal-to-noise ratio. The low absolute value for $Q$ is due to a large dispersion in $R_{\text{off}}$ and $R_{\text{on}}$ also observed in noiseless experiments as reported in, e.g., Ref.~\cite{Ghenzi.JournalAppliedPhysics.2012}. 

\begin{figure}
\begin{center}
\includegraphics[width=1\columnwidth]{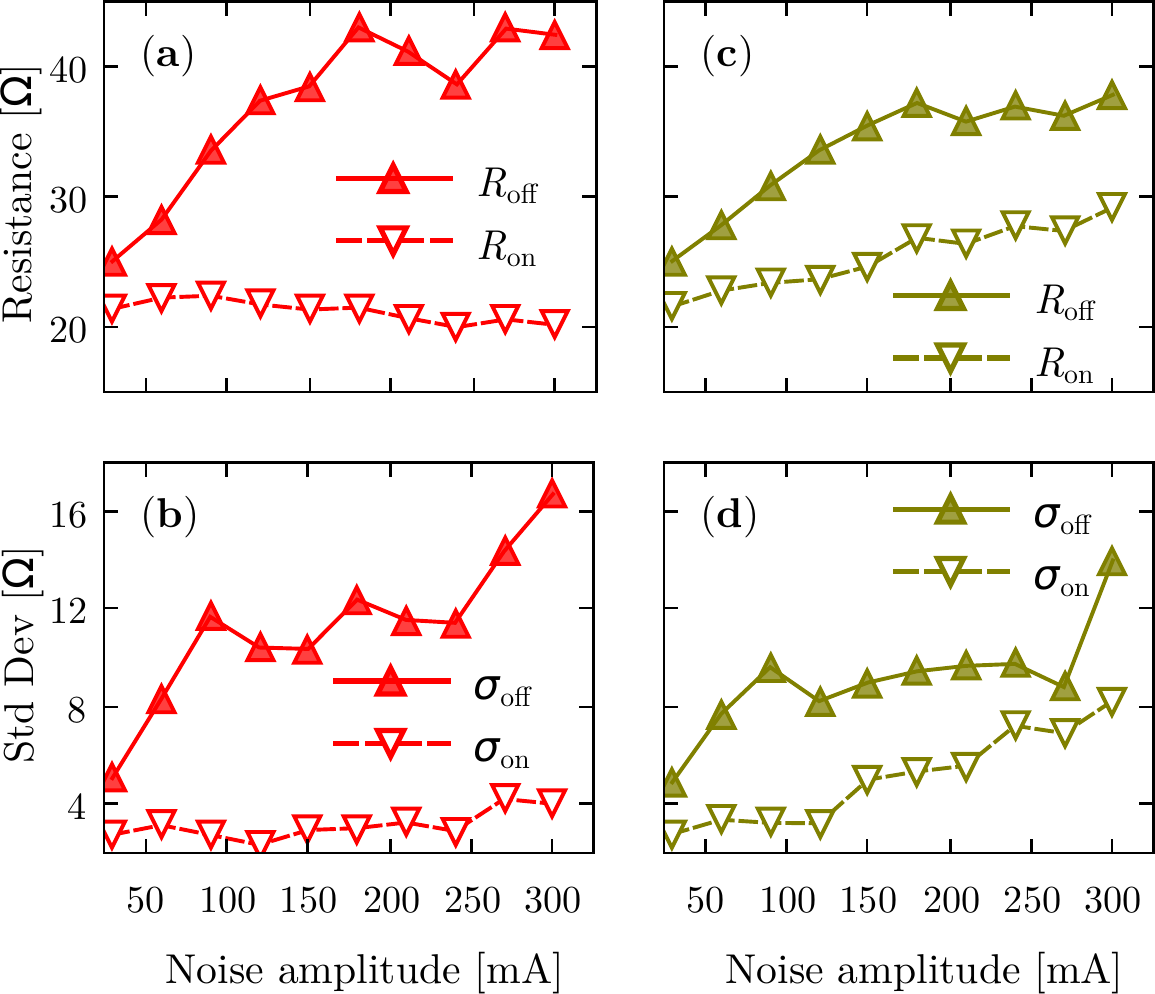}
\caption{(Color online). (a) and (c) show $\left\langle R_{\text{off}}\right\rangle$ and $\left\langle R_{\text{on}}\right\rangle$  \textit{vs}. noise amplitude.  (b) and (d) show $\sigma_{\text{off}}$ and $\sigma_{\text{on}}$. Figs. (a) and (b) show results 1 ms after the signal was turned off, while Figs.~\ref{fig:R_exp} (c) and (d) show results after 10 ms.}
\label{fig:R_exp}
\end{center}
\end{figure}

Initially, the addition of small noise helps increase the contrast between the $\left\langle R_{\text{off}}\right\rangle$ and $\left\langle R_{\text{on}}\right\rangle$ as it is shown in Fig.~\ref{fig:R_exp}a. Right after the input signal is turned off, noise starts to degrade this contrast (see Fig.~\ref{fig:R_exp}c). In fact, we found that for large elapsed times $\left\langle R_{\text{off}}(t)\right\rangle \approx \left\langle R_{\text{on}}(t)\right\rangle$. Figure~\ref{fig:Q_exp} shows a diminishing $Q$ factor with elapsed time.

As the noise amplitude increases, also $\sigma_{\text{off}}$ and $\sigma_{\text{on}}$ increase (see Figs.~\ref{fig:R_exp}b and \ref{fig:R_exp}d). The accretion of the standard deviation could have overridden the gain in $\left\langle \Delta R \right\rangle = \left\langle R_{\text{off}}\right\rangle - \left\langle R_{\text{on}}\right\rangle$ if it were not for the fact that $\sigma_{\text{on}}$ remains small even for large noise amplitudes. Results in Fig.~\ref{fig:R_exp}a show a small change of $\left\langle R_{\text{on}}\right\rangle$, which suggests that $R_{\text{on}}$ values reach a saturation point. This saturation, in turn, leads to a low dispersion of $R_{\text{on}}$ and a small $\sigma_{\text{on}}$.

\section{\label{sec:sim_res}Numerical results}

Encouraged by the agreement between the model by Pickett \textit{et al}. \cite{Pickett.JournalofAppliedPhysics.2009} and experimental results (see Ref.~\cite{Patterson.PhysRevE.87}), we performed a numerical study on the role of noise under different input conditions. Simulations were based on the simplified model by Kvatinsky \textit{et al}. \cite{Kvatinsky.IEEETransactionsonCircuitsandSystemsI.2013}, where the evolution of the state variable $x$ in Eq.~(\ref{eq:memristor_state}) is determined by the function 
\begin{equation}
f(x,i) = \begin{cases}
k_{\text{off}} \left( \frac{i}{i_{\text{off}}} - 1 \right)^{\alpha_{\text{off}}}f_{\text{off}}(x)  & 0 < i_{\text{off}} < i \\
0  & i_{\text{on}} < i < i_{\text{off}} \\
k_{\text{on}} \left( \frac{i}{i_{\text{on}}} - 1 \right)^{\alpha_{\text{on}}}f_{\text{on}}(x)  & i < i_{\text{on}} < 0 \ ,
\end{cases}
\label{eq:motion}
\end{equation}
where $i_{\text{on,off}}$, $k_{\text{on,off}}$, and $\alpha_{\text{on,off}}$ are parameters chosen to adjust Pickett's model. Functions $f_{\text{on,off}}(x)$ ensure that $x$ remains within $\left[x_{\text{on}}, x_{\text{off}} \right]$, and are given by
\begin{equation}
\begin{split}
f_{\text{off}}(x) &= \exp\left[ - \exp \left( + \frac{x - x_{\text{off}}}{w_{c}}  \right) \right]  \ ,\\
f_{\text{on}}(x) &= \exp\left[ - \exp \left( - \frac{x - x_{\text{on}}}{w_{c}}  \right) \right]  \ ,\\
\end{split}
\label{eq:window}
\end{equation}
where $w_{c}$ is a fitting parameter. Finally, resistance in Eq.~(\ref{eq:ohm}) is given by
\begin{equation}
R(x,i) =  R_{\text{a}}+R_{\text{b}} \exp{\left(\lambda \frac{x-x_{\text{on}}}{x_{\text{off}}-x_{\text{on}}} \right)} \ ,
\label{eq:ohm2}
\end{equation}
where $\lambda$,  $R_{\text{a}}$ and $R_{\text{b}}$ are suitably chosen constants.

\begin{figure}
\begin{center}
\includegraphics[width=1\columnwidth]{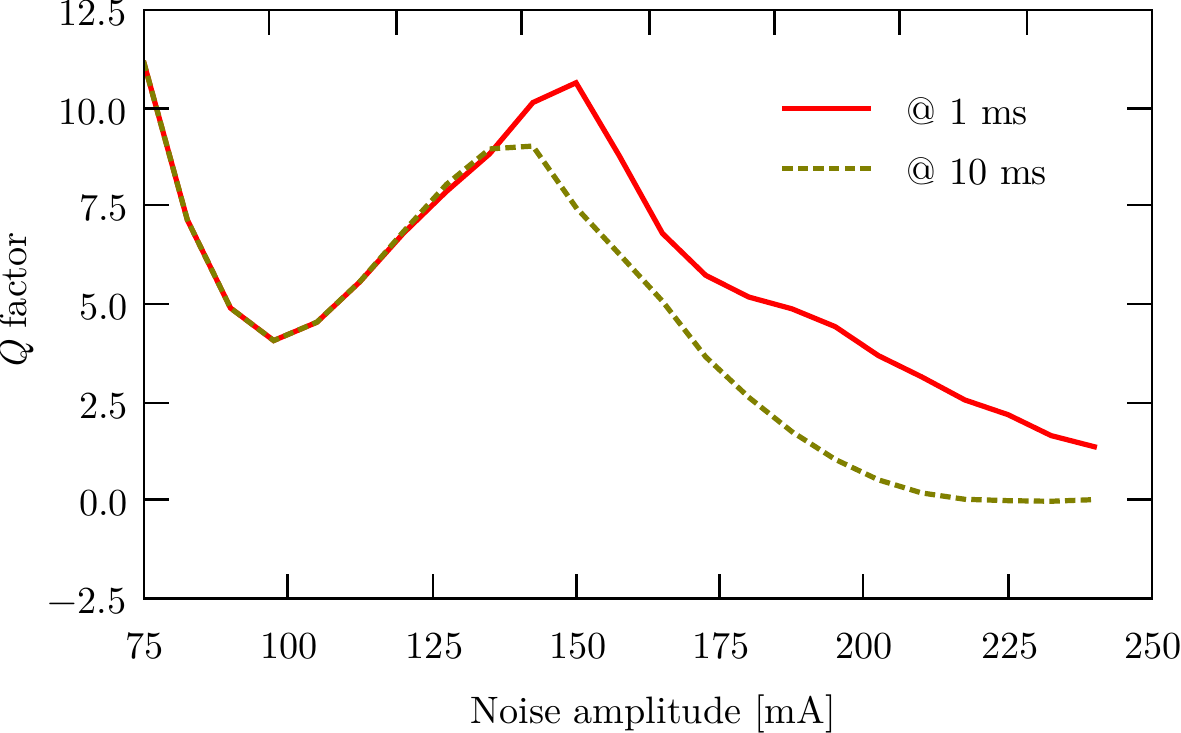}
\caption{(Color online). Quality factor as a function of noise amplitude. Results are shown for 1 and 10 ms after the input signal is turned off.}
\label{fig:Q_sim}
\end{center}
\end{figure}

We performed 10$^{3}$ noise realizations for each set/reset protocol and computed the $Q$ factor for noise amplitudes ranging from 40 to 240 mA. Fig.~\ref{fig:Q_sim} shows simulation results. As with the experimental results, a non-monotonic relation between the quality factor and the noise amplitude is found. However, the model fails to capture the fact that experiments with manganite samples suffer from low repeatability. This fact explains the apparent divergence of the $Q$ factor in the absence of noise in Fig.~\ref{fig:Q_sim}. Nevertheless, the simulated model captures the enhancement of the quality factor in the presence of a finite noise amplitude.

\begin{figure}
\begin{center}
\includegraphics[width=1\columnwidth]{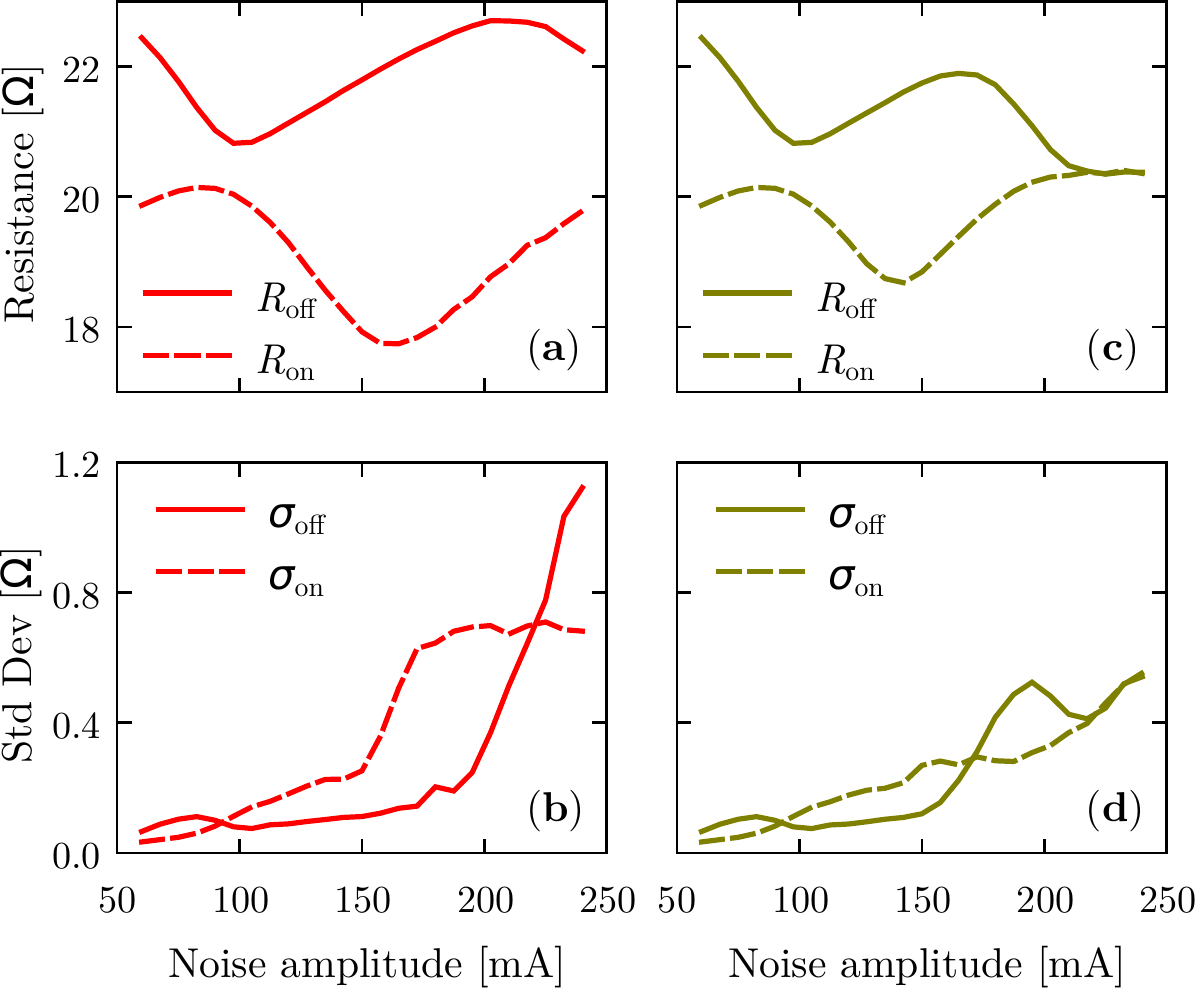}
\caption{(Color online). (a) and (c) show $\left\langle R_{\text{off}}\right\rangle$ and $\left\langle R_{\text{on}}\right\rangle$  as a function of noise amplitude.  (b) and (d) show $\sigma_{\text{off}}$ and $\sigma_{\text{on}}$. Figs. (a) and (b) show the results after 1 ms has elapsed since the signal was turned off while Figs.~\ref{fig:R_sim} (c) and (d) show results after 10 ms.}
\label{fig:R_sim}
\end{center}
\end{figure}

We believe that the beneficial role of noise can be explained by some general features of the model. Jensen's inequality states that $F \left( \left\langle \xi \right\rangle \right)$ $\leq$ ($\geq$) $\left\langle F\left( \xi \right) \right\rangle$, where $\xi$ is a random variable and $F$ is a convex (concave) function. In  Eq.~(\ref{eq:motion}), $f(x,i)$  is a convex (concave) function of $i$ for positive (negative) current. This implies that the net effect of a zero-mean random noise added to the input signal is to increase $\left\langle \dot x\right\rangle$ as compared to a noiseless input (see Eq.~(\ref{eq:memristor_state})). This, in turn, leads to a larger $\left\langle \Delta R\right\rangle$ (see Figs.~\ref{fig:R_sim}a and \ref{fig:R_sim}c). When the noise amplitude is too large, however, the input current changes sign, and its net effect in a single direction is canceled out. After the driving signal is turned off, noise eventually drives the resistance to a point that does not depend on the originally set state, that is, memory is lost, as it can be seen in Fig.~\ref{fig:R_sim}c. 

\begin{figure}
\begin{center}
\includegraphics[width=1\columnwidth]{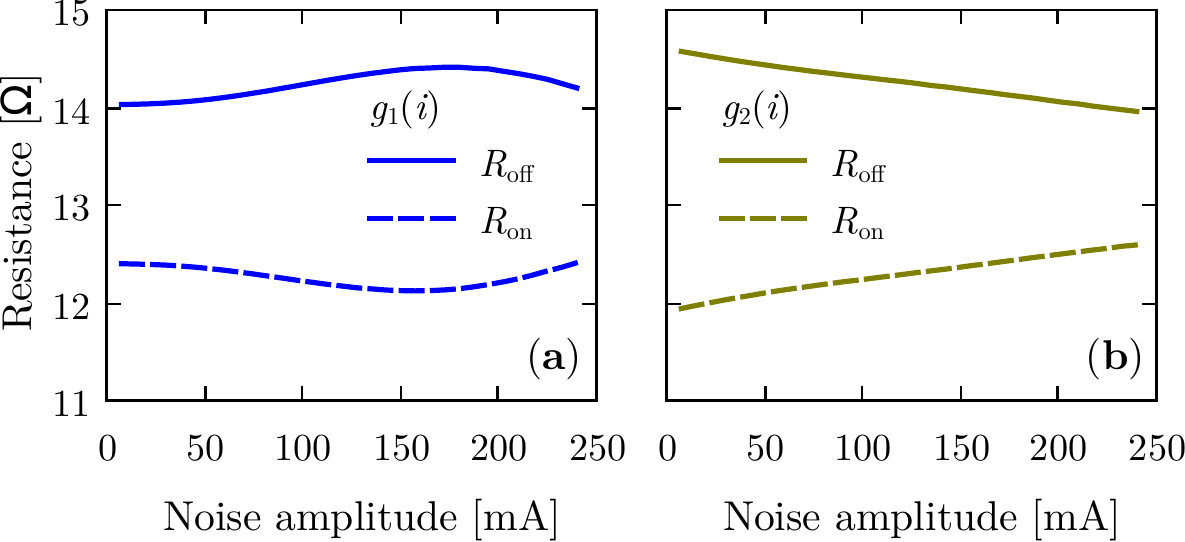}
\caption{(Color online). $\left\langle R_{\text{off}}\right\rangle$ and $\left\langle R_{\text{on}}\right\rangle$  as a function of noise amplitude. Fig. (a) shows results when considering $g_{1}(i)$  and (b) when $g_{2}(i)$ is taken into account. Results are computed after 1 ms has elapsed since the signal is turned off.}
\label{fig:R_sim2}
\end{center}
\end{figure}

To support this reasoning, we studied Eq.~(\ref{eq:memristor_state}) with
\begin{equation}
f(x,i) = \begin{cases}
g_{k}(i)f_{\text{off}}(x) & 0 < i\\
g_{k}(i)f_{\text{on}}(x) & i <  0 \ ,
\end{cases}
\label{eq:g_k}
\end{equation}
and two different functions
\begin{equation}
g_{1}(i) = c_1i^3\text{, and }g_{2}(i) = c_2 \tanh(c_3 i),
\label{eq:toymodels}
\end{equation}
where $c_i$ are arbitrarily fixed parameters. Note that we have eliminated any explicit thresholds like $i_{\text{off}}$ and $i_{\text{on}}$ in Eq.~(\ref{eq:motion}). Also note that $g_1$ is a simple convex (concave) function for positive (negative) current. On the contrary, $g_2$ is concave (convex) for positive (negative) current. Therefore, according to our explanation, a non-monotonic behavior of $\left\langle \Delta R\right\rangle$ as a function of noise should be observed only for $g_1$. Figs.~\ref{fig:R_sim2}a and \ref{fig:R_sim2}b show that, indeed, the role of noise is as predicted.

Figs.~\ref{fig:R_sim}b and \ref{fig:R_sim}d show that $\sigma_{\text{off}}$ and $\sigma_{\text{on}}$ increase with noise amplitude. As in the case of the experimental results, this increment could override the gain in $\left\langle \Delta R\right\rangle$ if it were not the case that the standard deviation corresponding to one of the resistive states remains small, even for a wide range of noise amplitudes. The small dispersion of $R_{\text{off}}$ is a saturation effect which is represented in the model by the extremely small values of the ``window function'' $f_{\text{off}}(x)$. The saturation in only one direction, in turn, emerges as a consequence of the asymmetry of the model for positive and negative currents. Such asymmetry accounts for the fact that the experimentally observed ON and OFF switching times are different~\cite{Strukov.Small.2009}.

\section{\label{sec:concl}Conclusions}

In this paper we studied the effect of external noise on resistive switching. We presented experimental results on a manganite sample that show a beneficial role of noise. In particular, we showed that a quality factor $Q$, which characterizes the contrast between high and low resistive states, is maximized for a finite noise amplitude.   

By means of numerical simulations based on a model in Ref.~\cite{Kvatinsky.IEEETransactionsonCircuitsandSystemsI.2013}, we studied the causes underlying the observed behavior. This analysis led us to relate experimental results to some general characteristics of the equations governing the dynamics of a memristor (Eqs.~(\ref{eq:ohm}-\ref{eq:memristor_state})). Local convexity (concavity) of $f(x,i)$, as a function of $i$, for positive (negative) current leads to an enhancement of the contrast between high and low resistive states when noise is added. Although this gain is accompanied by an increment in the dispersion of the resistive values, the latter is small for only one of the resistive states and for a wide range of noise intensities. The cause of this small dispersion is that the corresponding resistive state lies close to a saturation point which is, in turn, a consequence of the asymmetry in the ON and OFF switching times.

\acknowledgments The authors would like to thank  A.~G.~Leyva and P.~Levy from CAC-CNEA for providing the manganite sample. We gratefully acknowledge financial support from ANPCyT under project PICT-2010 No. 121.

\bibliography{biblio}

\end{document}